\documentclass[10pt,a4paper]{article}

\usepackage{graphicx}
\usepackage{dcolumn}
\usepackage{bm}
\usepackage{url}

\begin{document}

\title{Best linear unbiased estimation of the nuclear masses}

\author{Bertrand Bouriquet $^1$ \footnote{bertrand.bouriquet@cerfacs.fr}
   \and Jean-Philippe Argaud $^2$}

\maketitle

\footnotetext[1]{
Sciences de l'Univers au CERFACS, URA CERFACS/CNRS No~1875,
42 avenue Gaspard Coriolis,
F-31057 Toulouse Cedex 01 - France
}
\footnotetext[2]{
Electricit\'e de France,
1 avenue du G\'en\'eral de Gaulle,
F-92141 Clamart Cedex - France
}

\begin{abstract}
This paper presents methods to provide an optimal evaluation of the nuclear
masses. The techniques used for this purpose come from data assimilation 
that allows combining, in an optimal and consistent way, information coming from
experiment and from numerical model. Using all the available information, it
leads to improve not only masses evaluations, but also to decrease
uncertainties. Each newly evaluated mass value is associated with some accuracy
that is sensibly reduced with respect to the values given in tables, especially
in the case of the less well-known masses. In this paper, we first introduce a
useful tool of data assimilation, the Best Linear Unbiased Estimation (BLUE). This BLUE method
is applied to nuclear mass tables and some results of improvement are shown.

{\bf keyword:}  

Data assimilation, Best Linear Unbiased Estimation,  BLUE, nuclear masses, mass
tables

\end{abstract}

\section{Introduction}

The mass tables provide an evaluation of the mass for every known and forecast
nuclei and are very important in nuclear physic. Information gathered inside
those table by experimentalist and various nuclear mass models (for example
"Finite-Range Liquid-Drop Model" \cite{Wei1,Bet1} or "Finite-Range Droplet
Model" \cite{mol2}) is used for reaction planning and nuclear reactions
simulations. Thus, an accurate knowledge of masses of the nuclei permits to
realize high quality calculations and planning.

The purpose of this paper is to present a method to optimally evaluate masses of
known nuclei, as well as the accuracy associated. The aim is to produce an
improved set of data for nuclear masses, with better accuracy respect to
tabulated ones.  This general approach is already applied in other fields of
science, as for example in meteorology, or oceanography or neutronic
\cite{Bouriquet2010a,Bouriquet2010b}. The procedure proposed here is the same as
the one climatologists use to obtain high accuracy meteorological data. This is
the case for example of the widely used meteorological re-analysis ERA-40
\cite{era40}among others \cite{Kalnay96,Huffman97}.

Improving the accuracy of the data can be done in many ways, like in cumulating
information from various experiments in a cleaver way to reduce the global
inaccuracy, as presented in \cite{Rajput1992,Kafala1994,Macmahon2004}. Such
methods permit to converge toward a good accuracy of data. In the present case,
we are interested in including information coming from a numerical model to the
estimation of the value, and to the calculation of the associated accuracy. This
approach is reasonable as the model is giving an overall information on the
data, which means some constrains on the expected values for measurement. Data
assimilation is precisely a general method to handle jointly experimental data
and numerical modelling information to estimate the optimal values. Moreover,
data assimilation techniques allow at the same time to improve accuracy of the
estimation with respect to the original data.

In this paper, we will first develop some aspects on the theory and the basics
concepts of data assimilation. In fact data assimilation covers a large number
of techniques. Here we will focus on the Best Linear Unbiased Estimation (BLUE)
technique, that fits very well to the present problem. We will use this BLUE
technique to estimate the nuclear masses of the known nuclei found in the
classical mass tables. This will lead to a new set of nuclear masses with
improved accuracy.

\section{Data assimilation}\label{sec:da}

We briefly introduce the theory of  data assimilation. However, data
assimilation is a wide domain and we will not present here the advanced
techniques that include dynamics of the process, that are for example the basis
of the nowadays-meteorological operational forecast. Some interesting
information on these approaches can be found in the following references
\cite{tal97,kal03,cou99}. More recently some applications of data assimilation
have been done on nuclear core neutronic activity evaluation 
\cite{Massart07,Bouriquet2010a,Bouriquet2010b}

The ultimate goal of data assimilation methods is to be able to figure out the inaccessible
true value of the system state, so called $\mathbf{x}^t$ with the $t$ index for
"true". The basic idea of data assimilation is to put together information coming from an
\textit{a priori} on the state of the system (usually called $\mathbf{x}^b$,
with $b$ for "background"), and information coming from measurements (referenced
as $\mathbf{y}$).  The result of data assimilation is called the analysis $\mathbf{x}^a$, and
it is an estimation of the true state $\mathbf{x}^t$ we want to find. 

Some tools are necessary to achieve such a goal. As the mathematical space of
the background and the one of observations are not necessary the same, a bridge
between them needs to be built. This is the so called observation operator $H$,
with its linearisation  $\bf{H}$, that transforms values from the space of the
background to the space of observations. The reciprocal operator is the adjoint
of $H$, which in the linear case is the transpose $\bf{H}^T$ of $\bf{H}$.

Two other ingredients are necessary. The first one is the covariance matrix
$\bf{R}$ of observation errors, which are
$\epsilon_o=\mathbf{y}-H(\mathbf{x}^t)$. It can be obtained from the known
errors on the unbiased measurements. The second one is the covariance matrix
$\bf{B}$ of background errors, which are $\epsilon_b=
\mathbf{x}^b-\mathbf{x}^t$. It represents the error on the \textit{a priori},
assuming it to be unbiased. There are many ways to obtain these observation and
background error covariance matrices, such as kriging, stochastic simulation
algorithms, ensemble estimation, etc (see for example
\cite{Chiles1999,Goovaerts2001,Cheng2011}). However, this is commonly the output of a
model and an evaluation of its accuracy, or the result of expert knowledge. 

To find this optimal value $\mathbf{x}^a$ the underlying idea is to minimise the
variance of the error $\epsilon_a=\mathbf{x}^a-\mathbf{x}^t$ associated to this
value. Then it can be proved \cite{cou99} that, within this formalism, the Best Unbiased
Linear Estimator $\mathbf{x}^a$ is given by the following equation:
\begin{equation}\label{xa}
\mathbf{x}^a = \mathbf{x}^b + {\bf K} \big(\mathbf{y}-\mathbf{H}\mathbf{x}^b\big)
\end{equation}
where ${\bf K}$ is the gain matrix:
\begin{equation}\label{K}
{\bf K} = {\bf BH}^T ({\bf HBH}^T + \mathbf{R} )^{-1}
\end{equation}
Moreover we can obtain the analysis error covariance matrix $\bf{A}$,
characterising the analysis errors $\epsilon_a$. This matrix can be expressed
from $\bf{K}$ as:
\begin{equation}
\mathbf{A} = (\mathbf{I-KH})\mathbf{B}
\end{equation}
with $\mathbf{I}$ the identity matrix. Note that one way  to prove equation
\ref{K} is to minimize the trace of the matrix $\bf{A}$, leading also to prove
$\mathbf{x}^a$ is the optimal value we are looking for. The demonstration is
detailed in the reference \cite{cou99}. It can equivalently be proven through a
maximum likelihood hypothesis. This lead on overall to an improvement of the
accuracy, as it can be show in \cite{Rajput1992,Kafala1994,Macmahon2004}.

Then the main work is to evaluate as well as possible the observation operator 
${\bf H}$ and the two covariance matrices $\mathbf{B}$ and $\mathbf{R}$. We will
proceed with this task within the framework of the mass tables.

\section{Application of data assimilation to the nuclear mass evaluation}

To build the Best Linear Unbiased Estimation of the mass tables, we will work on
masses excess instead of masses themselves. The mass excess of a nucleus is the
difference between its actual mass and its mass number ($A \times u$) with $u$
the atomic mass unit, or "unified atomic mass" (see Table A in \cite{au2} for
information on this unit) and $A$ the number of nucleons.

For the background $\mathbf{x}^b$, we will take the reference data of mass
excess values proposed in \cite{mol1,mol1f} by P. Moller, J.R. Nix, W.D. Myers,
and W.J. Swiatecki, focusing on the masses obtained from the Finite-Range
Droplet Model \cite{mol2} with shell energy correction. Note the results are
very similar to the ones obtained from the Finite-Range Liquid-Drop Model
\cite{Wei1,Bet1}. In those theoretical tables, we will limit ourselves to the
known nuclei heavier than oxygen, that is with protons number $Z \ge 8$, which
is the lowest reliable value for the model.   


The experimental mass excess values are the one from G.~Audi and all reference
tables \cite{au1,au2,au3,au1f}. This remarquable collection of data also
includes the error associated to each measurement. This represents the
observation, denoted $\mathbf{y}$ in the previous section.

These two data series of mass excess values $\mathbf{x}^b$ and $\mathbf{y}$ are
in the same space. Then the required observation operator ${\bf H}$ is obviously
reduced to identity ${\bf I}$.

The covariance matrix $\mathbf{R}$ on observation errors is chosen to be a
diagonal matrix. On the diagonal, we put the known value of the uncertainties
given in experimental tables \cite{au1,au2,au3,au1f}.

The evaluation of the background errors covariance matrix $\mathbf{B}$ is
slighly more complicated. The first assumption is that background errors are
independent the one from the other. Then $\mathbf{B}$ reduces to a diagonal
matrix. As no information yet exists on the model accuracy, it is also assumed
as a second assumption that this accuracy is independent of each nuclei we
consider. Thus we only  need one global value of accuracy denoted as
$\sigma_b^2$. It remains to evaluate this unique value $\sigma_b^2$ on the
diagonal. For this purpose, we made a statistical study. The error between
$\mathbf{x}^b$ and  $\mathbf{x}^t$ is overestimated by the error between
$\mathbf{x}^b$ and ${\bf H}\mathbf{x}^t=\mathbf{y}$, that is globally more
variable. Thus, considering that, last relative error gives a penalty to the
accuracy on model evaluation. This is the third assumption we make to obtain
model error evaluation. To evaluate a value for the diagonal of ${\bf B}$, we
calculate the mean square of $\mathbf{x}^b-\mathbf{y}$ over all the available
nuclei. The mean square is $\sigma_b^2=0.652904$, that is
$\sigma_b=0.808024~MeV$. This is the value that we will put on the diagonal of
$\mathbf{B}$.

For informative purpose, it is interesting to look at the calculated mean value
of $\mathbf{x}^b-\mathbf{y}$, which it is equal to $-0.058326~MeV$. It is fairly
close to $0$, and then comfort the implicit hypothesis to be in a quasi-unbiased
case for background error estimation. It can be checked that the correlation
between the background error and the measurement error is $0.1998$. This low
value means that there are no significant correlation between those two error.
Such a condition is required to use data assimilation under good condition. 

\begin{center}
\begin{figure}
\includegraphics[width=\columnwidth]{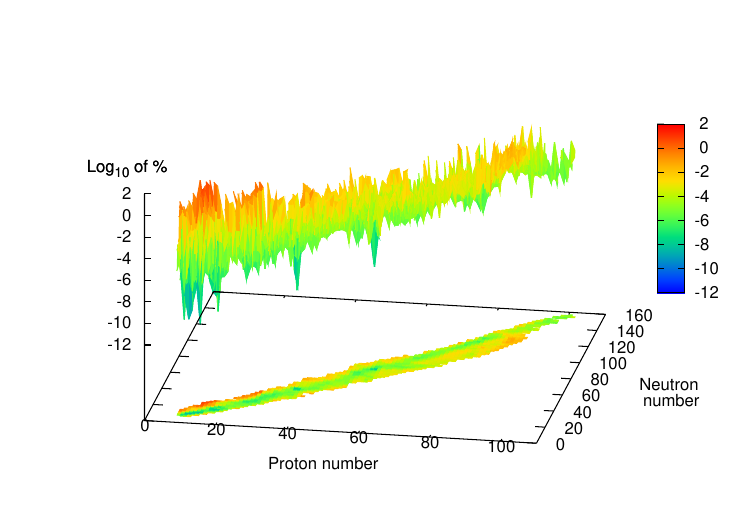}
\caption{Mass excess experimental values percentage of change, as a function of
the neutron and proton numbers of the nuclei.}
\label{fig1}
\end{figure}
\end{center}

All the required data are then available to build the data assimilation analysis
as described in section \ref{sec:da}, by simply applying formula \ref{xa}. The
differences between the experimental results $\mathbf{y}$ and the analysed
results $\mathbf{x}^a$ obtained with data assimilation are shown on Figure
\ref{fig1} in percentage of change of mass excess for all the nuclei in function
of their protons and neutrons numbers.  

On Figure \ref{fig1}, it can be noticed that the relative modification of the
masses could be as lower as $10^{-10}\%$ up to $80\%$ in absolute value. This
wide range of modification can be explained easily within an assimilation
process. On the one hand, in case nuclei which masses are known very accurately,
like for stable of nuclei, then information provided by the background (model) do
not contribute a lot, masses excess do not change, and the relative modification
is around $10^{-10}\%$. On the other hand, if nuclei masses excess are not known
very accurately masses excess given by the model give a lot more information.
Thus, information on physical property of nuclei included in the model allows to
drive the measured value toward a new value that is more likely (in the sense of
the maximum likelihood principle include in data assimilation method) and then
modification can be up to $80\%$. Thus, as we notice on Figure~\ref{fig1}, the
more we go far from stability valley the more modification of the nuclei masses
excess could be important because the less the experimental value are accurate
due to the extreme difficulty to realize such measurement.

Considering those notable modification it is worth taking into account the new
results for mass excess. We have checked that all results are  correctly
enclosed between the background previsions and the experimental values. This
mean that we never overshoot either experimental value or the one given by the
model. This result is in agreement with the modelling of ${\bf B}$ and  ${\bf R}$
that have been done.

A key point is on the accuracy, as, by construction of the method, data
assimilation improves it. The diagonal of the matrix ${\bf A}$ (which, in the
present case, is a diagonal matrix) contains the variance $\sigma_a^2$ of the
analysis $\mathbf{x}^a$ for each nuclear nuclei. We are looking at the
percentage of evolution of the accuracy, with respect to the experimental
accuracy $\sigma_y$ for each nuclei. Thus we can construct the following
accuracy indicator $\frac{\sigma_y-\sigma_a}{\sigma_y}$, observed in percent.

With such a definition, an improvement of the accuracy (that is a decrease of
$\sigma_a$ with respect to $\sigma_y$) is a positive percentage. 

\begin{center}
\begin{figure}
\includegraphics[width=\columnwidth]{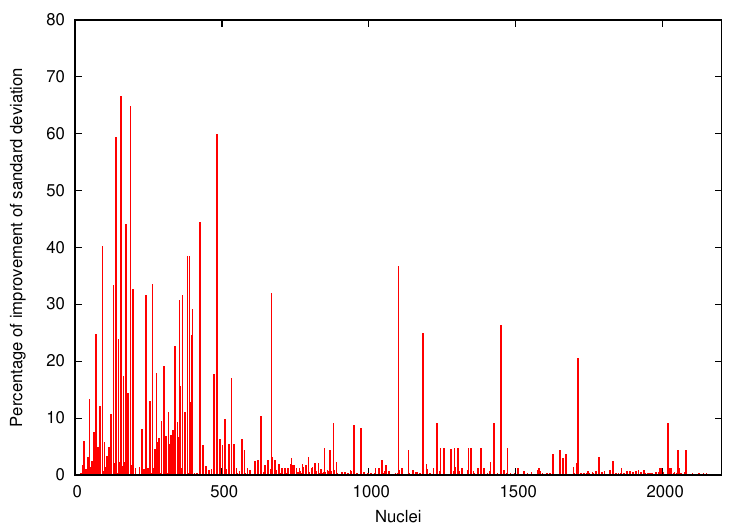}
\caption{Percentage of improvement of the experimental error by the analysed
error as a function of the number of the nuclei within the stables elements of
table \cite{au1f}.}
\label{fig2}
\end{figure}
\end{center}

We have done two representations of those values. The first one is an histogram
plot, where each bin represent the accuracy indicator for one nuclei, in order to
get a 1D representation of all the results. To obtain such mono-dimensional plot,
we consider the nuclei ordered in the same way as in the reference file 
\cite{au1f}. Thus, the first bin correspond to the first nuclei of the G. Audi
\textit{et al.} table, and so on. The results are presented on the
Figure~\ref{fig2}. This representation allows to have a global overview of the
amplitude of the modifications and of their values.  

\begin{center}
\begin{figure}
\includegraphics[width=\columnwidth]{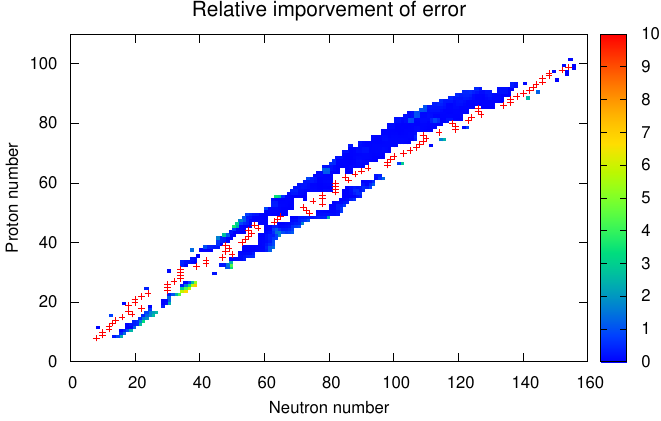}
\caption{Percentage of improvement of the experimental error respect to the
analysed error, as a function of the number of protons and neutrons. Each cross
represents the nuclei that is experimentally the most accurately measured for
each elements.}
\label{fig2-2D}
\end{figure}
\end{center}

A second representation, two-dimensional, shows the results of the same relative
accuracy as a function of the number of neutrons and protons. The results are
presented on the Figure \ref{fig2-2D} (for representation reason, the scale of
this figure is saturating at 10\%). On this one, the crosses represent the
nuclei that is the most accurately measured experimentally for each element.
This gives a line close to the stability valley. 

From Figure \ref{fig2}, we confirm that all values are positive, which means
that there is always an improvement of the accuracy. The improvement can be up
to roughly $70\%$ in some cases. As for the analysed value themselves, this
means that when an experimental value is known very accurately, a lot of efforts
are required to do better. On the contrary, if original accuracy on data is not
so good, it is easy to improve it, providing only little additional information.

Considering Figure \ref{fig2-2D}, from an experimental point of view, we notice
that data assimilation provide value with an improved accuracy for the elements
that are rather far from the stability line which usually mean the unstable
nuclei. Those nuclei are often measured with lower accuracy that the sable one
due to the intrinsic experimental difficulties.  

Then, globally speaking, we can say data assimilation method applied to nuclear
mass tables is very successful, and leads within a simple framework to some
significant improvements of nuclei masses excess and their associated error. 

\section{Conclusion}

Data assimilation technique applied on the mass tables seems then to be very
promising. Here is described the generation of an optimal set of masses that can
be used when needing mass tables information, as it was done in climatology by
ERA-40 re-analysis \cite{era40}. The new mass data set produced will prove to be
useful because:

\begin{itemize}
\item  it shows to be within the limit previously given by theory and
       experience,
\item  the accuracies on the mass excess are lower than the one previously
       known, which makes them more suitable to use.
\end{itemize} 

However, as a perspective, the present application is showing only some limited
aspects of the possibility of data assimilation. Especially, the improvement of
the $\mathbf{B}$ matrix can be studied in order to open the way for forecasting
more accurately masses of yet unknown nuclei.

\bibliographystyle{unsrt}
\bibliography{Nucphy}

\end{document}